\documentclass[10pt]{iopart}
\usepackage{iopams}
\expandafter\let\csname equation*\endcsname\relax
\expandafter\let\csname endequation*\endcsname\relax
\usepackage{amsmath}
\usepackage{graphicx}
\usepackage{physics}
\usepackage{bm,epsfig}
\usepackage[export]{adjustbox}
\usepackage{amssymb}
\usepackage{latexsym}
\usepackage{setspace}
\usepackage{color,soul}
\usepackage{cite}
\usepackage{gensymb}
\allowdisplaybreaks

\begin{document}
\title{Identifying orbital angular momentum of light in quantum wells}

\author{Seyedeh Hamideh Kazemi \& Mohammad Mahmoudi}

\address{Department of Physics, University of Zanjan, University Blvd., 45371-38791, Zanjan, Iran}
\ead{mahmoudi@znu.ac.ir}
\vspace{10pt}
\begin{indented}
\item[]November 2018
\end{indented}

\begin{abstract}
Generation and detection of structured light have recently been the subject of intense study, aiming to realize high-capacity optical storage and continuous-variable quantum technologies. Here, we present a scheme to extract the orbital angular momentum content of Laguerre-Gaussian light beams in a double-$\Lambda$ four level system of GaAs/AlGaAs multiple quantum wells. Arising from a quantum interference term, absorption of a non-vortex probe field depends upon the azimuthal phase of vortex fields so that both magnitude and sign of the azimuthal index/indices can be mapped into the absorption profile. 

\end{abstract}


\vspace{2pc}
\ioptwocol

\section{Introduction}
In 1992, Allen and co-workers \cite{Allen} demonstrated that a helically-phased light beam with azimuthal dependence of the form $\exp(i l \phi)$, with $l$ being as the azimuthal index, carries an intrinsic orbital angular momentum (OAM) content equal to per photon. The OAM, representing an extra degree of freedom for photons and a large state space to encode information, is exploited in quantum computation and optical communication \cite{Duan,Leach,Bozinovic}. Additionally, such large spaces and high-dimensional systems can, for instance, be proposed in order to increase information capacity as well as to improve the security of cryptographic keys \cite{Bourennane,Dada,Malik}. An example of such fields with a helical phase structure is the Laguerre-Gaussian (LG) light beam which has given birth to ground-breaking applications in quantum optics \cite{truscott,Das,kazemi1,kazemi,kazemi3}, optical micromanipulation \cite{He,Dholakia}, and quantum communication \cite{Terriza,giner}. 

Mechanisms for measuring LG modes were proposed in parallel to the development of tools to create such optical modes \cite{Dickey}. Of particular interest in this regard is measuring and identifying the OAM, mainly due to applications in communication, trapping and information processing. Techniques to identify the OAM have been based on  interferometric sorting \cite{Leach2,Lavery}, diffraction \cite{Berkhout,Hickmann}, optical transformation
through refractive elements \cite{Berkhout2,Dudley2}, and quantum interference phenomenon \cite{Han,Radwell,Hamedi}. A particular instance of these techniques is a scheme based on spatially-varying optical transparency; by measuring transmission of a vortex light beam in an atomic medium, an azimuthal modulation of the absorption profile- induced by phase and polarization structure of the beam- was observed \cite{Radwell}. 

On the other hand, a great deal of attention has recently been devoted to semiconductor quantum wells (SQWs), owing to the similarity between them and atomic vapors, and several interesting phenomena of atomic-molecular systems, e.g. optical coherence and interference effects, have been extended to SQWs. Some of these intersting phenomena are gain without inversion \cite{Imamoglu}, electromagnetically induced transparency (EIT) \cite{Sadeghi,Phillips}, enhanced index of refraction \cite{Sadeghi2,wang3,kang}, optical soliton \cite{solitons}, and spatial distribution of probe absorption \cite{sahrai}, to name but a few. In addition to these schemes, the relative phase of applied fields is considered as a method for controlling quantum coherent and interference in atomic \cite{Buckle,Korsunsky,Evers}, molecular \cite{Shapiro} and solid-state systems \cite{Kurizki,Joshi}. The advantages of using SQWs are their large electric dipole moments of intersubband transitions, high nonlinear optical coefficients, great flexibility in device design, and that their transition energies and dipole moments as well as symmetries can be also constructed with a high degree of accuracy from selected materials. 

In this Letter, we propose a scheme to identify LG modes in a double-$\Lambda$ four level system of GaAs/AlGaAs multiple quantum wells (MQWs) and demonstrate how absorption profile of a non-vortex probe field depends upon both sign and magnitude of the azimuthal index/indices, due to the closed-loop structure of the scheme. This work is motivated by a work of Hamedi \textit{et al}, \cite{Hamedi}, however, the main advantages of our suggested scheme are: 1) The probe absorption in their work do not provide any additional information about the sign of OAM, while absorption profile in this work can reveal both magnitude and sign of azimuthal index/indices. 2) Our approach to identify the characteristics of OAM is based on a coherently driven semiconductor quantum well nanostructure, which is easier to apply and thus more practical than that its gaseous counterpart; the medium studied here provides a highly tunable quantum system with its flexible design in such a way that its properties such as the transition energies and dipole moments can be engineered as desired by accurately tailoring its shape and size. 3) Moreover, Hamedi and coworkers focused on azimuthal modulation of EIT in a highly-resonant five-level combined tripod and $\Lambda$ atom-light coupling setup, which requires cold and trapped atoms. While in our suggested mechanism based on the MQW medium, similar effects can be found in ambient temperature conditions.  Thus, the ease-of-use of this suggested scheme can simplify a possible implementation of high-capacity data storage technologies and quantum information.

\section{Model and equations}

\begin{figure}[t]
\centering
\includegraphics[width=7cm]{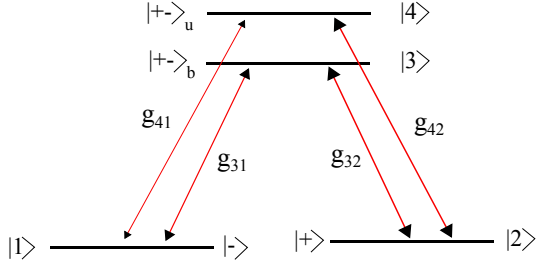}
\caption{ The figure illustrates the energy-level diagram of a four-level double-$\Lambda$ system. Exciton states are labeled in ascending energy: two lower states $\vert 1\rangle$ and $\vert 2\rangle$ plus two excited states $\vert 3\rangle$ and $\vert 4\rangle$. Two upper levels correspond to unbound and bound two-exciton states $\vert +- \rangle_{b}$, and  $\vert +- \rangle_{u}$, whereas two exciton states $\vert - \rangle$ and $\vert + \rangle$ are chosen to be lower states. }
\label{fig1}
\end{figure}

As shown in figure~\ref{fig1}, we consider a four-level double-$\Lambda$ system in a semiconductor MQW, consisting of twenty periods of wells which are assumed to be grown on a GaAs substrate. Each period consists of two GaAs wells, which separated by a 20 nm AlGaAs barrier. We denote two exciton states $\vert - \rangle$ and $\vert + \rangle$, by $\vert 1\rangle$ and $\vert 2\rangle$, respectively, while two upper levels, i.e., $\vert 3\rangle$ and $\vert 4\rangle$, correspond, respectively, to unbound and bound two-exciton states ($\vert +- \rangle_{b}$, and  $\vert +- \rangle_{u}$). The system interacts with four coherent fields; the ground level $\vert 2\rangle$ is coupled to excited levels $\vert 3\rangle$ and $\vert 4\rangle$ by two strong control fields, $E_{32}$ and $E_{42}$, with carrier frequencies $\omega_{32}$ and $\omega_{42}$, respectively. A weak probe field ($E_{41}$) with frequency $\omega_{p}=\omega_{41}$ couples $\vert 1\rangle$ and $  \vert 4\rangle$, while the transition $\vert 1\rangle - \vert 3\rangle$ is driven by a weak coupling field ($E_{31}$) with frequency $\omega_{31}$. It may be noted that such structures have been already studied for investigating slow light and tunable amplification in Refs~\cite{Ma} and \cite{Yan}. 

Under electric-dipole, rotating-wave approximations and in a suitable reference frame, the Hamiltonian is written as 
\begin{align}
 H^{I} &=  \hbar (\Delta_{32}-\Delta_{31} ) \tilde{\rho}_{22} - \hbar \Delta_{31} \tilde{\rho}_{33}  \\ \nonumber
&+  \hbar (\Delta_{32}-\Delta_{31} - \Delta_{42}) \tilde{\rho}_{44} \\ \nonumber
&-\hbar ( \, g_{31} \tilde{\rho}_{31} + g_{32} \tilde{\rho}_{32} +g_{42} \tilde{\rho}_{42} + g_{41} \tilde{\rho}_{41} e^{-i \Phi} + \mathrm{H.c.} ).
\label{eq1}
\end{align}
Here, H.c. corresponds to the Hermitian conjugate of terms in the Hamiltonian and $\Delta_{ij}= \omega_{ij}-\bar{\omega}_{ij}$ is the detuning of the laser field from corresponding transition, with $\bar{\omega}_{ij}$ being as the frequency of the transition $ \vert i\rangle \leftrightarrow \vert j\rangle$ ($i \in \lbrace 3,4 \rbrace$ and $j \in \lbrace 1,2 \rbrace$). Further, Rabi frequencies for the relevant laser-driven transitions can be written as $g_{ij}=(\vec{\mu}_{ij}. \vec{E}_{ij})/{\hbar}$, where $\vec{\mu}_{ij}$ and $\vec{E}_{ij}$ are the dipole moment of the corresponding transition and peak amplitude of the field, respectively. Additionally, we have defined $\rho_{mn} = \vert m\rangle \langle n \vert$ and the parameter $\tilde{\rho}_{mn}\,$ $(m,n$ $ \in \lbrace1, ..., 4\rbrace)$ denotes the corresponding operator in the new reference frame. As is clear from equation (1), the time dependence appears only in parameter $\Phi $, which is given by 
\begin{subequations}
\begin{align}
\Phi &= \Delta t- \vec{K} \vec{r} +\phi_0,  \\
\Delta &= (\Delta_{32} +\Delta_{41})-(\Delta_{31} +\Delta_{42}),  \\
\vec{K} &= (\vec{k}_{32}+\vec{k}_{41})-(\vec{k}_{31}+\vec{k}_{42}), \\
\phi_{0} &= (\phi_{32}+\phi_{41})-(\phi_{31}+\phi_{42}),
\end{align}
\end{subequations}
where $\Delta$, $\vec{K}$, and $\phi_{0} $ are the multiphoton resonance detuning, wavevector mismatch, and initial phase difference, respectively. Also, $\vec{k}_{ij}$ and $\phi_{ij}$, respectively, denote the wavevector and the absolute phase of each field. Noting that in writing the Hamiltonian, we have assumed that frequencies fulfill $\omega_{41}+\omega_{32}=\omega_{42}+\omega_{31}$, which will be reduced to $\Delta=0$ in this closed system. This condition also implies $k_{41}+k_{32}=k_{42}+k_{31}$ for close values of the laser frequencies, fulfilling the criteria of wave-vector mismatch: $\vec{K} = 0$. Thus, the parameter $\Phi $ reduces to a constant relative phase of applied phase $\phi_{0}$.

On the other hand, LG beam (LG$_{p}^{l}$) defines a solution of the paraxial wave function in a cylindrical coordinate with  $l$ and $p$ being as azimuthal and radial indices. The azimuthal index (sometimes called topological charge or winding number of light), characterizing phase dependence around the optical axis, are the number of times the phase completes 2$\pi$ on a closed loop around the axis of propagation and $p+1$ denotes the number of bright high-intensity rings around the beam propagation axis \cite{hanle}. 

Denoting coordinates as (x,y)=($r,\varphi$) in either Cartesian or polar, the Rabi frequency for LG beams (LG$_{0}^{l}$) is written as  
\begin{equation}
g_{ij} (r,\varphi)=  g^{'}_{ij} \,(\dfrac{ \sqrt{2}r}{w})^{l} \,e^{ - (r/w)^2 -i l_{ij} \varphi}=  u_{ij}(r) e^{-i l_{ij} \varphi}.
\end{equation}
Here, we have further defined $g^{'}_{ij}$, $\exp (-i l_{ij} \varphi)$, and the real function of $u_{ij}(r)$ as Rabi frequency constant, azimuthal phase, and mode amplitude, respectively. The parameter $w$ represents the width of the beam and is equal to $\mathrm{w} \sqrt{1+z^2/z^{2}_{R}}$ with $\mathrm{w}$ being as the beam waist. The $z$ dependence ca be ignored in the region $z\ll \,$ $z_{R}$, where $z_{R}$=$\pi \mathrm{w}^{2}/ \lambda$ is the Rayleigh range of the beam \cite{allen2}. Note that, for simplicity, we have assumed $p = 0$, allowing only for doughnut modes of order $l$, however, the treatment can, in principle, be applied also to any other LG beams.

The dynamic of the probe field can be described by density-matrix equations of motion which are written as follows
\begin{subequations}
\begin{align}
\frac{d}{dt}  \tilde{\rho}_{11} &= i g_{31}^{*} \tilde{\rho}_{31}- i g_{31} \tilde{\rho}_{13}+ i g_{41}^{*} \tilde{\rho}_{41} e^{i\Phi}  \\ \nonumber
& - i g_{41} \tilde{\rho}_{14} e^{-i \Phi} +2 \gamma_{4} \tilde{\rho}_{44}+ 2 \gamma_{3} \tilde{\rho}_{33},\\
\frac{d}{dt}  \tilde{\rho}_{22} &= i g_{32}^{*} \tilde{\rho}_{32}- i g_{32} \tilde{\rho}_{23}+ i g_{42}^{*} \tilde{\rho}_{42} - i g_{42} \tilde{\rho}_{24}  \\ \nonumber
&-  2 \gamma_{2} \tilde{\rho}_{22},\\
\frac{d}{dt}  \tilde{\rho}_{33} &=- i g_{31}^{*} \tilde{\rho}_{31}+ i g_{31} \tilde{\rho}_{13}- i g_{32}^{*} \tilde{\rho}_{32} + i g_{32} \tilde{\rho}_{23}  \\ \nonumber
&- 2 \gamma_{3} \tilde{\rho}_{33},\\
\frac{d}{dt}  \tilde{\rho}_{12}&= i ( \Delta_{32}-\Delta_{31} )\tilde{\rho}_{12} + i g_{31}^{*} \tilde{\rho}_{32}+ i g_{41}^{*} \tilde{\rho}_{42}e^{i \Phi}  \\ \nonumber
&- i g_{32} \tilde{\rho}_{13}-i g_{42} \tilde{\rho}_{14}- \gamma_{2} \tilde{\rho}_{12},\\
\frac{d}{dt}  \tilde{\rho}_{13}&=- i \Delta_{31} \tilde{\rho}_{13} + i g_{31}^{*}( \tilde{\rho}_{33}- \tilde{\rho}_{11})- i g_{32}^{*} \tilde{\rho}_{12} \\ \nonumber
&+ i g_{41}^{*} \tilde{\rho}_{43} e^{i\Phi}- \gamma_{3} \tilde{\rho}_{13},\\
\frac{d}{dt}  \tilde{\rho}_{14}\ &= i(\Delta_{32}-\Delta_{31}-\Delta_{42}) \tilde{\rho}_{14}+ i g_{31}^{*} \tilde{\rho}_{34}  \\ \nonumber
 &+ i g_{41}^{*} e^{i\Phi} ( \tilde{\rho}_{44}- \tilde{\rho}_{11})- i g_{42}^{*}  \tilde{\rho}_{12}- \gamma_{4} \tilde{\rho}_{14},\\
\frac{d}{dt}  \tilde{\rho}_{23}\ &=- i \Delta_{32} \tilde{\rho}_{23} + i g_{32}^{*}( \tilde{\rho}_{33}- \tilde{\rho}_{22})- i g_{31}^{*} \tilde{\rho}_{21} \\ \nonumber
 &+ i g_{42}^{*} \tilde{\rho}_{43} -(\gamma_{2}+\gamma_3) \tilde{\rho}_{23}, \\
\frac{d}{dt}  \tilde{\rho}_{24}\ &=- i \Delta_{42} \tilde{\rho}_{24} + i g_{42}^{*}( \tilde{\rho}_{44}- \tilde{\rho}_{22})+i g_{32}^{*} \tilde{\rho}_{34}  \\ \nonumber
&- i g_{41}^{*} \tilde{\rho}_{21} e^{i\Phi}-(\gamma_{2}+\gamma_4) \tilde{\rho}_{24},\\
\frac{d}{dt}  \tilde{\rho}_{34}\ &=- i(\Delta_{42}-\Delta_{32}) \tilde{\rho}_{34} + i g_{31}  \tilde{\rho}_{14} +i g_{32} \tilde{\rho}_{24} \\ \nonumber
&- i g_{41}^{*} \tilde{\rho}_{31} e^{i\Phi} - i g_{42}^{*} \tilde{\rho}_{32}- (\gamma_{3}+\gamma_4) \tilde{\rho}_{34}.
\label{equ3} 
\end{align}
\end{subequations}
The remaining equations follow from $\tilde{\rho}_{mn}=\tilde{\rho}^{*}_{nm}$ and trace condition $\sum _{m} \tilde{\rho}_{mm}=1 $. The decay constants for dipole-allowed transitions are denoted by $\gamma_{2}$, $ \gamma_{3}$, and $ \gamma_{4}$, which are the sum of population decay rates and dephasing ones. The former are due to longitudinal optical phonon emission events at low temperature and the latter are determined by electron-phonon scattering processes and interface roughness \cite{Tsujino}. The following values for decay constants are predicted for a typical quantum well structure: $ \gamma_{3}=  \gamma_{4}=$2-3 meV and $ \gamma_{2}=  0.8$ meV \cite{Joshi}. Here, we reiterate that coefficients of equations (4) do not have an explicit time dependence under multiphoton resonance conditions, $\Delta=0$ and $K=0$, and one can find a stationary steady-state in the long-time limit.

On the other hand, the linear susceptibility of the weak probe field can be written as $\chi = N \eta_{p} \varrho_{41}/(\epsilon_0 E_{41})$ with $N$, $\eta_{p}$, and $\varrho_{41}$ being as, respectively, density of carriers, probe transition dipole moment, and probe coherence. Notice that the coherence appeared in this relation refers to the one oscillating at a frequency of the incident probe field, i.e., $\e^{i \alpha_{41} } \rho_{41} \equiv \e^{i\Phi } \tilde{\rho}_{41}$, with $\alpha_{41}=(\omega_{p} t -\vec{k}_{41} \vec{r} +\phi_{41})$. By setting $N \eta_{p}^2 /( \hbar\epsilon_0) =1$, the susceptibility now reads $\chi = \e^{i\Phi } \tilde{\rho}_{41}/ g_{41}$ \cite{Evers}. As is well known, real and imaginary parts of the susceptibility correspond to dispersion and absorption, respectively. Moreover, in our notation, positive (negative) values in the imaginary part of susceptibility correspond absorption (gain) for the probe field.

In order to calculate the probe absorption, we need to obtain the coherence term related to the probe field, $\tilde{\rho}_{41}$, from equations (4). Under the weak-probe field approximation, in which coupling and control fields are much stronger than the probe one, it can be assumed that almost all excitons are in the ground state and the expression of the probe absorption can be written as:
\begin{equation}
\mathrm{Im}[\chi] = -\dfrac{A \gamma  (2 \gamma \gamma^{'} - g_{31}^{2} + g_{32}^{2} +     g_{42}^{2}   )}{B g_{41}   } .
\end{equation}
Here, $\gamma_3$=$\gamma_4$=$\gamma^{'}$, $\gamma$=$\gamma^{'}$+$\gamma_2$, $A$=$\mathrm{Im}[i\, \e^{i \phi_{0}} g_{31} g_{42} g_{32}^{*} ] $, and $B$ =$\gamma \gamma^{'} g_{31}^{4} + g_{31}^{2}(\gamma \gamma^{'} (2 \gamma^{'2} + \gamma \gamma_2 ) + g_{32}^{2} (\gamma^{'2} + \gamma^{2}) )   + \gamma \gamma^{'} ( 2 \gamma^{'} \gamma_2 \gamma^2 + \gamma^{'} (2 \gamma + \gamma_2) g_{32}^{2} +   g_{32}^{4} + \gamma^{'} (\gamma_2-2 \gamma) g_{42}^{2} -g_{42}^{4}  )  $. Generally, the analytical expression for steady-state of $\tilde{\rho}_{41}$ can be derived up to higher-order in $g_{41}$, however, our calculation shows that this zero-order solution yields the main contribution for the scheme. 

The term $A$, which is called the "quantum interference term", corresponds to a closed-interaction loop and scattering of  driving field modes into the probe field one, as it contains the product of three driving fields ($g_{31} g_{42} g_{32}^{*}$). This round-trip sequence of dipole-allowed transitions embodies a prominent point: When one or a combination of three driving fields carry a vortex, it is, in principle, possible to obtain information about the OAM of those fields from the absorption profile.

\section{Results and Discussions}

In the following, we present solutions of the probe absorption, which are obtained by numerically solving equations~(4) in the long-time limit, for a variety of conditions and demonstrate how the absorption is dependent on both the magnitude and the sign of the azimuthal index/indices associated with one or a combination of driving fields. Starting with the case of only one field with an LG profile, we investigate the effect of its azimuthal index as well as the relative phase of applied fields on the probe absorption. In particular, we show that the absorption profile can reveal the sign of the azimuthal index of the optical vortex, in addition to detect the modulus. We then continue with three different configurations in which two of driving fields have LG profiles: (1) two control fields, $E_{42}$ and $E_{32}$, are vortex beams, (2) $E_{31}$ and $E_{32}$ carry optical vortices and (3) when $E_{42}$ and $E_{31}$ are assumed to be vortex ones. Finally, we consider the case that three driving fields, $E_{31}$, $E_{32}$, and $E_{42}$, have LG profiles. It is worth mentioning that, throughout the discussion, we have concentrated on situation which the probe field does not have any optical vortex.

For a realistic example, one may consider the parameters of the GaAs/AlGaAs quantum wells to numerically investigate the probe absorption: $\gamma^{'}$= $2\times 10^{12}$Hz and $ \gamma_{2}$=$0.8 \times 10^{12}$Hz \cite{Yan}. We also assume that carrier frequencies of fields satisfy the multi-photon resonance condition, i.e., $\Delta_{31}$ = $\Delta_{32}$ = $\Delta_{42}$ = 0 \cite{Phillips}. Moreover, our results are represented in scaled quantities to obtain the best possible comparison with other similar schemes; positions are divided by $\mathrm{w}$ with a typical value of the beam waist $ 10\, \mu$m \cite{Schmiegelow}. We have also used dimensionless Rabi frequency (field amplitude) and Rabi frequency constants: $g_{ij}/\gamma^{'}$ and $g_{ij}^{'}/\gamma^{'}$, respectively.

\begin{figure}[t]
\centering
\includegraphics[width=7cm]{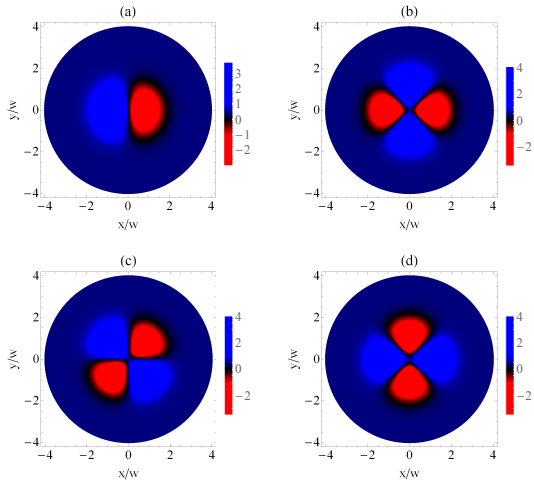}
\caption{ The probe absorption versus normalized positions ($x/\mathrm{w}$,$\,y/\mathrm{w}$) in the case of a vortex control beam $g_{42}$ and for (a) $l_{42}=1$, (b) $l_{42}=2$ with $\Phi=0$, whereas (c) and (d) depict profiles for $l_{42}=2$ with $\Phi=\pi/2$ (c) and $\Phi=\pi$ (d). The common parameters are chosen as $ \gamma_{3}=  \gamma_{4}= 2\times 10^{12}$Hz, $ \gamma_{2}= 0.8 \times 10^{12}$Hz, $\Delta_{31} = \Delta_{32} = \Delta_{42} = 0$, $\mathrm{w}= 10\, \mu$m, $g^{'}_{42}= g_{32}= 1 $, $g_{31}= 0.1  $, and $g_{41}= 0.01 $.  }
\label{fig2}
\end{figure}

First of all, we consider the case that one of the control fields ($g_{42}$) has an LG profile, but other driving fields have no vortices. Considering $\Phi=0$ and the Rabi frequency of the vortex field as $u_{42}(r) e^{-i l_{42} \varphi}$, quantum interference term appeared in equation (5) would be $g_{31} g_{32}^{*} u_{42}(r) \, \cos(l_{42} \varphi)$. For brevity, we will omit hereafter Rabi frequency of non-vortex fields and radial dependence of vortex ones in such terms and only consider Sine/Cosine functions, due to their corresponding azimuthal phase factor, as quantum interference terms. There are two important points to notice about this term. First, for incident field with azimuthal index $l_{42}= l $, the absorption profile is exactly the same as that of a negative one ($-l$). Also, this term suggest that the probe absorption varies cosinusoidally with a periodicity of $l$. In figure~\ref{fig2}, we show numerical results of the spatially-dependent absorption profile for azimuthal index $l=1$ in figure~\ref{fig2}(a) and $l=2$ in figure~\ref{fig2}(b). The parameters are $\Delta_{31}$ = $\Delta_{32}$ = $\Delta_{42}$ =$\Delta^{'}$= 0, $g^{'}_{42}=g_{32}$= 1, $g_{31}$= 0.1, and $g_{41}$= 0.01. Red areas in each plot indicate the regions of gain in the spectra (the negative imaginary part of the absorption), while blue and black ones represent positions of large absorption and transparency, respectively. As is clear form figures~\ref{fig2}(a) and \ref{fig2}(b), the profile shows a $l$-fold symmetry so that by counting the number of red (blue) lobes, the unknown vorticity of the LG beam can be easily recognized. The above results can be generalized to higher azimuthal indices with the same general trend: the number of lobes appeared on corresponding patterns would be equal to $2l$.

We then proceed to investigate the effects of the relative phase of applied fields on the profile. Based on the equation (5), one would expect that changing the relative phase may rotate the absorption profile; for instance, the quantum interference term is changed to $-\sin (l \varphi)$ by increasing the relative phase to $\pi/2$. The appearance of Sine function in this case, suggests that absorption profile can also act as a detector of sign of the azimuthal index; by reversing sign of the index, peaks (dips) in the absorption profile convert to dips (peaks), so that appeared pattern in the profile rotates by 90$\degree$.

\begin{figure}[t]
\centering
\includegraphics[width=7cm]{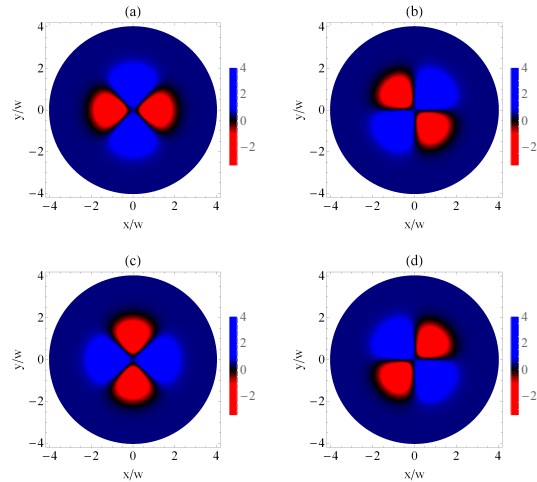}
\caption{ Spatially-dependent absorption profile for different relative phase of applied fields and $l_{42}=-2$. The phase difference between four laser fields is $\Phi=0$ in (a), $\Phi=\pi/2$ in (b), $\Phi=\pi$ in (c), and $\Phi=3\pi/2$ in (d). Other parameters are the same as for figure~\ref{fig2}(b). }
\label{fig3}
\end{figure}
These points and more are clearly seen from figures~\ref{fig2}(c) and \ref{fig2}(d) in which calculated profiles are shown for $\Phi=\pi/2$ and $\pi$, respectively. The other parameters are kept the same as in figure~\ref{fig2}(b). By increasing the relative phase of applied fields, patterns in the absorption profile undergo an anti-clockwise rotation; for instance, pattern for $\Phi=\pi/2$ exhibits an rotation of -$45 \degree$ (see figures~\ref{fig2}(b) and \ref{fig2}(c)). In the case of $\Phi=\pi$, and for a constant azimuthal number, the quantum interference term would be proportional to $-\cos(l \varphi)$, thus, the pattern rotates by -90$\degree$ for $\Phi=\pi$, as shown in figure~\ref{fig2}(d). By this we mean that, in patterns that have formed in this spatially-dependent absorption profile, peaks switch to dips and vice versa, so that patterns remain, but with an anti-clockwise rotation of $90 \degree$. These figures also show that similar results hold for nonzero relative phase of applied fields so that absorption profiles display expected $ \vert 2 l \vert$ lobes. It is worth mentioning that very similar results have been obtained for two other configuration in which either $g_{32}$ or $g_{31}$ has an LG profile, thus we have omitted such figures for the sake of space. Furthermore, detailed calculations show these results are independent of choosing detunings so that a similar trend is found for the case of nonzero detunings, but with peaks (dips) of smaller height (depth) for $\Delta^{'} > \gamma^{'}$. From what has been just discussed, one can indeed ascertain the rule that the number of lobes on appeared patterns is twice the modulus of the azimuthal index associated with one of the applied fields.

\begin{figure*}[!ht]
\centering
\includegraphics[width=15cm]{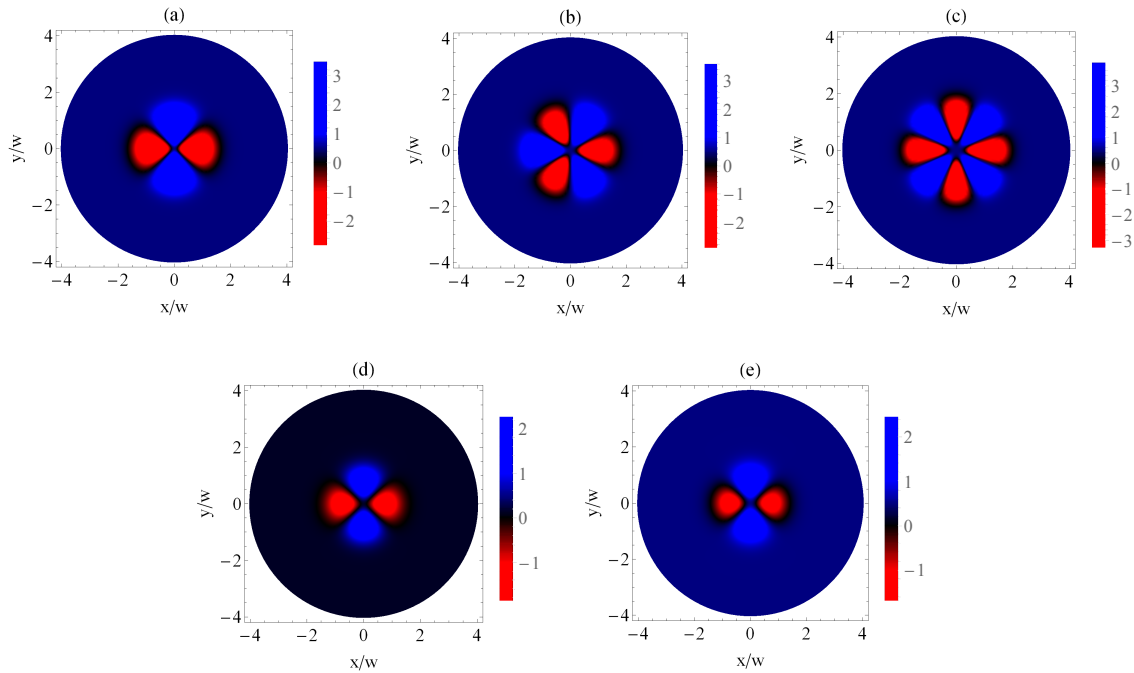}
\caption{ The figure shows spatially-dependent absorption profiles, when two of driving fields have optical vortices. Plot (a), (b), and (c) are for the case of two control fields, $E_{42}$ and $E_{32}$ with indices ($l_{42}$,$l_{32}$)=(1,-1), (1,-2), and (2,-2), respectively. Rabi frequencies are chosen as $g^{'}_{42}=g^{'}_{32}$= 1. The profile also displays a four-lobed pattern when $E_{31}$ and $E_{32}$ have LG profiles with ($l_{31}$,$l_{32}$)=(1,-1) and  $g^{'}_{32}$= 1, $g^{'}_{31}$= 0.1 in (d). A similar pattern is also seen for the case of $E_{42}$ and $E_{31}$ with vorticities of ($l_{42}$,$l_{31}$)=($1$,$1$) and $g^{'}_{42}$= 1 and $g^{'}_{31}$= 0.1 in (f). Other parameters are the same as for figure~\ref{fig2}(b). }
\label{fig4}
\end{figure*}

So far, we have considered only positive azimuthal index, but the present method is applicable to detect identification of vortex beams with negative one. Figure~\ref{fig3}(a) depicts the spatially-dependent absorption profile in the same unit, when the sign of azimuthal index associated with the control field is changed to negative; The other parameters are kept the same as in figure~\ref{fig2}(b). As is shown, in the case of $\Phi=0$, the profile does not contain any information about sign of the azimuthal index, as the pattern is the same for both positive and negative indices (see figures~\ref{fig2}(b) and \ref{fig3}(a)). However, sorting out positive and negative modes of LG fields can be possible by changing the relative phase of applied fields, as discussed before. Figure~\ref{fig3}(b) shows the absorption profile for a negative index of $l=-2$ and $\Phi=\pi/2$. As deduced from the quantum interference term for this case- $ \sin ( l \varphi)$ which is the reverse of the corresponding term for positive index- the absorption profile is changed, pointing out that rotating direction of the pattern can reveal information about sign of the azimuthal index; in other words, a $90 \degree$ rotation of pattern in the absoprtion profile indicates that the control field carries an opposite index. In this context, it would be imperative to mention the work of Han and coworkers \cite{Han} in which a scheme, based on the EIT modulated by a microwave field in atomic ensembles, is adapted to sort out negative and positive modes of an LG beam by moving the position of the atomic cell.

Figure~\ref{fig3}(c), on the other hand, shows the absorption profile for same parameters in figure~\ref{fig2}(d) and $l=-2$. By increasing the relative phase to $\pi$, the pattern undergoes an clockwise rotation of $90 \degree$. Another point to mention here is that there is a general trend as the same one displayed in figure~\ref{fig3}(a); the sign of the azimuthal index can not be distinguished by the corresponding pattern. As a matter of fact, a simple derivation can prove that the sign of indices can not be recognized by absorption patterns for $\Phi = n \pi$, with $n$ being as an integer, since patterns are exactly the same for positive and negative indices. The absorption profile for $\Phi=3 \pi/2$ is shown in figure~\ref{fig3}(d), which clearly shows expected rotation. It is needless to say that the pattern would be fully rotated back to its initial condition, when the phase switched to $\Phi=2 \pi$. By comparing figures~\ref{fig2} and~\ref{fig3}, it is clear that the rotating direction of the pattern is related to sign of indices; by increasing the relative phase from $0$ to $2 \pi$, the absorption profile for the positive (negative) azimuthal indices undergoes an anti-clockwise (clockwise) rotation, as can be verified by inspecting of equation (5). 

According to above results, one may conclude that the number of lobes and rotation of the pattern can be useful as a detector of the azimuthal index of vortex fields. What is noticeable, and which deserves to be highlighted, is that such an identification is obviuosly different from schemes on transfer of orbital angular momentum of light which were worked out in atomic ensembles \cite{Ruseckas,Ruseckas2,Amini}. Here, the probe beam does not acquire any vortices from driving fields with a well-defined OAM; instead, whose absorption profile can reflect azimuthal phase variation of fields with LG modes, mainly due to obtain some OAM components from such fields.

We then consider configurations in which two of driving fields have LG profiles. First, we investigate the situation that two control fields, $g_{32}$ and $g_{42}$, are vortex beams; figures~\ref{fig4}(a)-(c) show numerical results of spatially-dependent absorption profile with indices ($l_{42}$,$l_{32}$)=($1$,$-1$), (1,-2), and (2,-2), respectively. Also, Rabi frequencies are chosen as $g^{'}_{42}=g^{'}_{32}$= 1 and other parameters are kept the same as in figure~\ref{fig2}(b). In this case, quantum interference term becomes proportional to $ \cos ( (l_{32} -l_{42}) \varphi)$ with $l_{32}$ and $l_{42}$ being as indices of control fields. As is clear from this term, the absorption profile shows a $2l$-fold symmetry for the case of the opposite helicity vortex beams $l_{32}=-l_{42}=l$ (figures \ref{fig4}(a) and (c)). For different vortices, $ \vert l_{32} \vert \neq \vert l_{42} \vert$, profile reveals a $2 \vert l_{32} -l_{42} \vert$-lobed pattern, which is consistent with what is expected from equation (5); the probe absorption varies cosinusoidally with a periodicity of $  \vert l_{32} - l_{42} \vert$, as illustrated in figure \ref{fig4}(b). It may be of interest to note that a similar term to that obtained in this work, i.e., $\mathrm{Im}[i\, \e^{i \phi_{0}} g_{42} g_{32}^{*} ] $ was the subject of another investigation in an atomic system, as has recently reported by the present authors and their colleague\cite{local}. 
\begin{figure}[t]
\centering
\includegraphics[width=7cm]{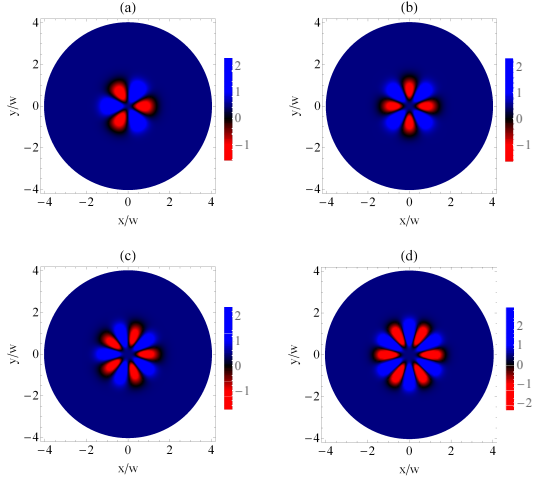}
\caption{ Spatially-dependent absorption profile when three driving fields, $E_{31}, E_{32}$ and $E_{42}$, have optical vortices; ($l_{31}$,$l_{32}$,$l_{42}$)=(1,-1,1) in (a), (1,-2,1) in (b), (1,-2,2) in (c), and (2,-2,2) in (d). Rabi frequencies are $g^{'}_{32}$=$g^{'}_{42}$=1, $g^{'}_{31}$=0.1, and other parameters are the same as for figure~\ref{fig2}(b).}
\label{fig5}
\end{figure}
It was found that utilizing two control vortex beams with equal azimuthal indices, $l_{42}$=$l_{32}$, can permit ultra-high precision and spatial resolution atom localization. Under this assumption, the scheme becomes independent of the azimuthal phase and reported features are mainly attributed to radial dependence associated with LG beams.

In the second configuration, driving fields $g_{31}$ and $g_{32}$ are assumed to carry vortices. Figure~\ref{fig4}(d) shows the absorption profile for indices ($l_{31}$,$l_{32}$)=(1,-1), $g^{'}_{32}$=$g_{42}$=1, $g^{'}_{31}$= 0.1, and with other parameters same as those in figure~\ref{fig2}(b). As is shown in this figure, the  profile shows two-fold symmetry, satisfying the cosinusoidal dependency predicted in equation (5), to be more exact $ \cos ( (l_{32} -l_{31}) \varphi)$. We then assume that driving fields $g_{31}$ and $g_{42}$ carry optical vortices. In figure~\ref{fig4}(f), we show the corresponding profile for vorticities of ($l_{42}$,$l_{31}$)=(1,1) and $g^{'}_{31}$= 0.1. Other parameters are the same as those in figure \ref{fig2}(b). The absorption profile in this figure reflects azimuthal indices associated with these fields; the profile clearly displays expected four lobes, as quantum interference term takes the form of $ \cos ( (l_{31} + l_{42}) \varphi)$. 


\begin{figure}[t]
\centering
\includegraphics[width=7.4cm]{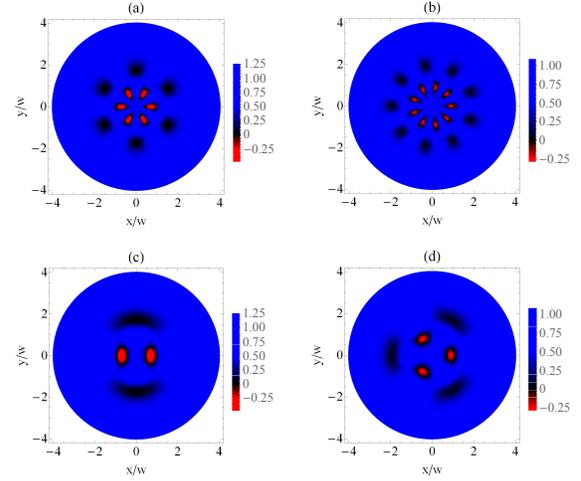}
\caption{ Spatially-dependent absorption profile for three LG beams with $LG_{p=1}^{l}$ modes. In the first row, we have assumed that $E_{31}$ and $E_{42}$ have equal indices, $l_{31}$=$l_{42}$=$l$ opposite to the index of another vortex beam $l_{32}$. (a) and (b) show profiles for $l=2$ and $l=3$, respectively. The second row correspond to $l_{31}$=$l_{42}$=$l_{32}$=$l$ with $l=2$ in (c) and $l=3$ in (d). Other parameters are the same as for figure~\ref{fig5}(a).}
\label{fig6}
\end{figure}
 
We then consider three driving fields $g_{31}, g_{32}$ and $g_{42}$ as vortex beams and plot corresponding profiles for vorticities ($l_{31}$,$l_{32}$,$l_{42}$)=(1,-1,1) in (a), (1,-2,1) in (b), (1,-2,2) in (c), and (2,-2,2) in (d). Rabi frequencies are $g^{'}_{31}$=0.1, $g^{'}_{32}$=$g^{'}_{42}$=1 and other parameters are taken as the same in figure~\ref{fig2}(b). In this case the quantum interference term reads as $ \cos ( (l_{31}+l_{42}- l_{32}) \varphi)$. When $E_{31}$ and $E_{42}$ have equal indices ($l_{31}$=$l_{42}$=l), opposite to the index of the another vortex beam $l_{32}$, the term will reduce to $ \cos ( 3l \varphi)$ and the profile will show a $3l-$fold symmetry (see figures~\ref{fig5}(a) and \ref{fig5}(d)). While for different indices, its symmetry is changed to $\vert l_{31} +l_{42}- l_{32} \vert$-fold, arising from a Cosine function appeared in equation (5) with a periodicity of $\vert l_{32} +l_{42}- l_{32} \vert$. Resulting absorption profiles are shown in figures~\ref{fig5}(b) and \ref{fig5}(c) which reveal a 8-lobed and 10-lobed, respectively.

Although beyond this work's scope, to demonstrate the ability to generalize the analysis to $LG_{p}^{l}$ modes, we demonstrate spatially-dependent absorption profiles for vortex beams with nonzero radial index modes whose Rabi frequencies are given by
\begin{align}
& G_{ij} (r,\varphi) =  g^{'}_{ij} \, (\dfrac{ \sqrt{2}r}{\mathrm{w}})^{l} \,L_{p}^{\vert l \vert} [\dfrac{2r^2}{\mathrm{w}^2}] \,e^{ - (r/\mathrm{w})^2 -i l_{ij} \varphi},
\end{align}
where $L_{p}^{\vert l \vert}$ represents associated Laguerre polynomial. Noting that, here we restrict ourselves to $p$=1, just for simplicity and for the sake of space, and calculate absorption profiles of a non-vortex probe field for the case of three driving beams with $LG_{1}^{l}$ modes. In the first row of figure~\ref{fig6}, it has been assumed that $E_{31}$ and $E_{42}$ have equal indices ($l_{31}$=$l_{42}$=$l$), opposite to the index of the another vortex beam $l_{32}$. Figures~\ref{fig6}(a) and \ref{fig6}(b) depict corresponding profiles for $l$=2 and $l$=3, respectively. Apart from two rings in these figures due to the radial modes of LG beams, their corresponding pattern consists of $6 \vert l \vert$ lobes in each ring which are spaced equally on a circle centered with the beam axis. In fact, by recalling equation (5) and $L_{1}^{\vert l \vert}(X)= 1+ \vert l \vert -X $, one can readily find the quantum interference term for such modes as $ \cos( 3 l \varphi)$. In addition, the distribution of absorption (gain) structure in those profiles is complementary between two rings. The second row in this figure correspond to $l_{31}$=$l_{42}$=$l_{32}$=$l$ with $l=2$ in figure~\ref{fig6}(c) and $l=3$ in figure~\ref{fig6}(d). Under this situation, the quantum interference in equation (5) takes the form of $\cos (l \varphi)$, resulting in $2l$-lobed patterns in absorption profiles. More specifically, each ring in these pattern has specific red (blue) areas which can be a sign of azimuthal indices of driving fields, similar to what we have seen in previous cases.

\section{Conclusions}

In conclusion, this paper has proposed and analyzed a novel scheme for the detection of information stored in the OAM carried by driving fields, which takes advantage of interesting features in the closed-interaction loop. Under the multi-photon resonance conditions and in the double-$\Lambda$ four level system of GaAs/AlGaAs MQW, we have demonstrated how profile absorption of the non-vortex probe field reflects the azimuthal phase variation of driving fields with LG modes. We have shown that one can obtain information about the modulus of the index/indices associated with one or a combination of three driving fields, through measuring the probe absorption spectra. Most prominently, due to the closed-loop structure of the scheme, sorting positive and negative modes of LG fields can be also possible. 

\section{Acknowledgments}
This work is supported by the Iran National Science Foundation (Grant No. 96008805).

\end{document}